\documentclass{sig-alternate-05-2015}

\begin{document}

\newcommand{\citex}{\textsc{Citex}}
\newcommand{\cA}{\mathcal{A}}
\newcommand{\cP}{\mathcal{P}}
\newcommand{\cV}{\mathcal{V}}
\newcommand{\xx}{\mathbf{x}}
\newcommand{\yy}{\mathbf{y}}
\newcommand{\zz}{\mathbf{z}}

\setcopyright{acmcopyright}

\doi{10.475/123_4}

\isbn{123-4567-24-567/08/06}

\conferenceinfo{International Workshop on
Mining and Learning with Graphs,}{KDD 2016, San Francisco, California, USA}

\acmPrice{\$15.00}

%

\title{A Graph Analytics Framework for Ranking Authors, Papers and Venues}

\numberofauthors{2} 
%
\author{
\alignauthor Arindam Pal\\
       \affaddr{TCS Research}\\
       \affaddr{Tata Consultancy Services}\\
       \affaddr{Kolkata -- 700156, India}\\
       \email{arindam.pal1@tcs.com}
\alignauthor Sushmita Ruj\\
       \affaddr{Computer and System Sciences Division}\\
       \affaddr{Indian Statistical Institute}\\
       \affaddr{Kolkata -- 700108, India}\\
       \email{sush@isical.ac.in}
}

\date{May 24, 2016}

\maketitle
\begin{abstract}
A lot of scientific works are published in different areas of science, technology, engineering and mathematics. It is not easy, even for experts, to judge the quality of authors, papers and venues (conferences and journals). An objective measure to assign scores to these entities and to rank them is very useful. Although, several metrics and indexes have been proposed earlier, they suffer from various problems. In this paper, we propose a graph-based analytics framework to assign scores and to rank authors, papers and venues. Our algorithm considers only the link structures of the underlying graphs. It does not take into account other aspects, such as the associated texts and the reputation of these entities. In the limit of large number of iterations, the solution of the iterative equations gives the unique entity scores. This framework can be easily extended to other interdependent networks.
\end{abstract}

%
%

%
%
\printccsdesc



\section{Introduction}
An \emph{author} is a person who publishes papers in some venues. A \emph{paper} is a scientific publication, such as research/technical papers, books and book chapters. A \emph{venue} is a place for publishing papers, like conferences, journals and workshops.

There are some general characteristics of good authors, papers and venues, which intuitively make sense. We list them below.
\subsection{Characteristics of a Good Author}
\begin{itemize}
	\item Should have independent research capabilities.
	\item Should have many publications.
	\item Should have \emph{important, high-quality} publications.
	\item Should have lot of citations for his papers.
	\item Should publish in \emph{good} venues.
	\item Should work with other \emph{good} authors.
\end{itemize}

\subsection{Characteristics of a Good Paper}
\begin{itemize}
	\item Should be written by \emph{good} authors.
	\item Should have many citations.
	\item Should be cited by other \emph{good} papers.
	\item Should be published in \emph{good} venues.
\end{itemize}

\subsection{Characteristics of a Good Venue}
\begin{itemize}
	\item \emph{Good} authors should publish there.
	\item \emph{Good} papers should be published there.
	\item Should be cited by \emph{good} papers published in other \emph{good} venues.
\end{itemize}

There is no \emph{perfect} citation index. Every index has some problems. A researcher can manipulate an index to his own advantage. We will explain some of the issues below.
\begin{enumerate}
	\item \textbf{Self-citation:} An author can give citation to her own papers to boost her number of citations. Sometimes it's necessary, sometimes it is dishonesty. It can be easily detected.
	\item \textbf{Mutual citation:} Researcher $A$ gives citation to a paper of researcher $B$. Researcher $B$ gives citation to a paper of researcher $A$. This is a mutual citation of length 2. It can be a longer cycle also. Now some people may form a cycle or even a clique of size $k$, and cite each others papers, thereby increasing their citation counts unfairly. This can be easily detected by a graph-based fraud detection algorithm.
	\item \textbf{Big shots and the rich get richer phenomenon:} Suppose someone is a \emph{big shot}, like a professor who has many PhD students. She is the center of a star graph. Any publication from her group will have her name, no matter whether she actually worked on that paper. Her paper and citation count will increase unfairly. This can be easily detected. This is also called the \emph{rich get richer phenomenon}.
	\item \textbf{Working in a large group:} Suppose a large group of ten people works on some problems. Any paper that comes out of that group has names of all the researchers. That increases the number of papers and citations artificially.
	\item \textbf{Type of publication:} If you publish a book, book chapter or survey paper, that will have many more citations compared to a research paper. That does not mean that the book/survey has more research output.
	\item \textbf{Area of research:} Some areas will have more citations than others. Papers in Natural Sciences like Physics or Biology have more citations than Computer Science. Within Computer Science, Machine Learning papers have more citations than Theoretical Computer Science papers.
	\item \textbf{Difficulty of papers:} Generally, the more difficult and mathematical a paper is, the lesser the number of citations, because less people will understand and cite the paper. That does not mean that the paper is less important.
	\item \textbf{Conference/journal type:} Some conferences typically have far more citations than others. ICASSP or ICC will have many more citations than STOC or KDD, because the first 2 accepts 1000+ papers, while the last 2 accepts 50-100 papers. That does not mean STOC/KDD is less important than ICASSP/ICC. In fact, people generally agree that it is much harder to get a paper accepted in STOC/KDD than ICASSP/ICC. This is also related to (6) and (7).
	\item \textbf{Division of credits:} Some people get credit just because they are in a research group without giving any significant ideas. It is difficult to detect this.
\end{enumerate}

\section{Related Works}
Many earlier works tried to rank authors and papers based on their importance. Some improtant ones are Hirsch's $h$-index \cite{H05}, Individual $h$-index \cite{BCK06}, Egghe's $g$-index \cite{E06}, and Zhang's $e$-index \cite{Z09}. One of the earliest work in this area is by Garfield \cite{G55}, who defined a metric called \emph{impact factor}. The impact factor of a journal is the average number of citations received by papers published in the  journal in the past two years. Pinski and Narin \cite{PN76} proposed a citation index by observing that not all citations are equally important. They noted that a journal is \emph{important}, if it is cited by other \emph{important} journals. Kleinberg \cite{K99} gave the first algorithm to rank web pages based on the hyperlink structure of the web graph. This algorithm is called Hypertext Induced Topic Search (HITS). 

Brin and Page came up with the \textsc{PageRank} citation metric \cite{brin1998anatomy,page1999pagerank} to use with the Google web search engine. This algorithm uses a \emph{random walk} model on the web graph. They proved that the \textsc{PageRank} is the \emph{stationary distribution} of this random walk. Gy{\"o}ngyi et al. \cite{gyongyi2004combating} gave an algorithm to detect spam web pages, which they called \textsc{TrustRank}. The \textsc{SimRank} algorithm by Jeh and Widom \cite{jeh2002simrank} gives a measure of the similarity between two objects based on their relationships with other objects. Their basic idea is that two objects are similar if they are related to similar objects. Zhou et. al. \cite{zhou2007co} proposed a method for co-ranking authors and their publications using several networks associated with authors and papers. Their co-ranking framework is based on coupling two random walks that separately rank authors and documents using the \textsc{PageRank} algorithm.

Walker et. al. \cite{walker2007ranking} gave a new algorithm called \textsc{CiteRank}. The ranking of papers is based on a network traffic model, which uses a variation of the \textsc{PageRank} algorithm. Chen et. al. \cite{chen2007finding} used a \textsc{PageRank} based algorithm to assess the relative importance of all publications. Their goal is to find some exceptional papers or \emph{gems} that are universally familiar to physicists. Sun and Giles \cite{sun2007popularity} proposed a popularity weighted ranking algorithm for academic digital libraries. They use the popularity of a publication venue and compare their method with the \textsc{PageRank} algorithm, citation counts and the HITS algorithm. Pal and Ruj \cite{PalR15} designed the {\citex} citation index to assign scores to both authors and papers. They considered the structures of the publication graph and the citation graph.

\section{Problem Definition and Model}
\label{sec:Problem definition}
\subsection{The Fundamental Graphs}
\begin{itemize}
	\item \textbf{Collaboration graph:} This graph is associated with the authors. The nodes of the graph are the authors. There is an undirected edge between two nodes, if the corresponding authors have written a paper together.
	\item \textbf{Citation graph:} This graph is associated with the papers. The nodes of the graph are the papers. There is a directed edge from a paper to another paper, if the first paper has cited the second paper.
	\item \textbf{Publication graph:} This graph is relating the authors, papers and venues. The nodes of the graph are the authors, papers and venues. There is an undirected edge between an author node and a paper node, if the author has written the paper. There is an edge between an author and a venue, if the author has published in that venue. There is an edge between a paper and a venue, if the paper was published in that venue.
\end{itemize}

We have a set of $m$ \emph{authors} $\cA = \{a_1,\ldots,a_m\}$, a set of $n$ \emph{papers} $\cP = \{p_1,\ldots,p_n\}$, and a set of $r$ \emph{venues} $\cV = \{v_1,\ldots,v_r\}$. A venue can either be a \emph{journal}, \emph{conference} or \emph{workshop}. We represent this by a \emph{publication graph} $G_P = (V_P, E_P)$, whose vertices are the set of authors, papers and venues, \emph{i.e.}, $V_P = \cA \cup \cP \cup \cV$. There are three types of edges in this graph.

\begin{itemize}
	\item There is an undirected edge between author $a_i$ and paper $p_j$, if author $a_i$ has written paper $p_j$. Note that, this is a \emph{symmetric} relation, so the edges are \emph{undirected}. Associated with this, there is an $m \times n$ \emph{author-paper matrix} $M$, whose rows and columns are $a_1,\ldots,a_m$ and $p_1,\ldots,p_n$ respectively, and whose $(i,j)^{th}$ entry $m_{ij} = 1$, if and only if author $a_i$ has written paper $p_j$.
	\item There is an undirected edge between author $a_i$ and venue $v_k$, if author $a_i$ has published a paper at venue $v_k$. Associated with this, there is an $m \times r$ \emph{author-venue matrix} $N$, whose rows and columns are $a_1,\ldots,a_m$ and $v_1,\ldots,v_r$ respectively, and whose $(i,k)^{th}$ entry $n_{ik} = 1$, if and only if author $a_i$ has published a paper at venue $v_k$.
	\item There is an undirected edge between paper $p_j$ and venue $v_k$, if paper $p_j$ is published at venue $v_k$. Associated with this, there is an $n \times r$ \emph{paper-venue matrix} $L$, whose rows and columns are $p_1,\ldots,p_n$ and $v_1,\ldots,v_r$ respectively, and whose $(j,k)^{th}$ entry $l_{jk} = 1$, if and only if paper $p_j$ is published at venue $v_k$.
\end{itemize}
Since there are only edges between authors, papers and venues, the publication graph is a \emph{tripartite graph}.

There is a \emph{collaboration graph} $G_A = (V_A, E_A)$ associated with the authors, whose vertices are the set of authors, \emph{i.e.}, $V_A = \cA$. There is an \emph{undirected} edge between author $a_i$ and author $a_j$, if they have written a paper together. Note that collaboration (co-authorship) is not a \emph{transitive} relation. Suppose authors $a_i$ and $a_j$ have written a paper $p_r$ and authors $a_j$ and $a_k$ have written a paper $p_s$. It may be that $a_i$ and $a_k$ have not written a paper togeher. Associated with this, there is an $m \times m$ \emph{collaboration matrix} $Q$, whose both rows and columns are $a_1,\ldots,a_m$, and whose $(i,j)^{th}$ entry $q_{ij} = 1$, if and only if author $a_i$ has written a paper with author $a_j$.

There is a \emph{citation graph} $G_C = (V_C, E_C)$ associated with the papers, whose vertices are the set of papers, \emph{i.e.}, $V_C = \cP$. There is a \emph{directed} edge from paper $p_j$ to paper $p_k$, if paper $p_j$ has cited paper $p_k$. Note that this is an \emph{asymmetric} relation, so the edges are \emph{directed}. Associated with this, there is an $n \times n$ \emph{citation matrix} $C$, whose both rows and columns are $p_1,\ldots,p_n$, and whose $(j,k)^{th}$ entry $c_{jk} = 1$, if and only if paper $p_j$ has cited paper $p_k$. Note that the citation graph can't have any \emph{directed cycle}. This is because a paper can only cite a previously published paper, so they are \emph{totally ordered} in time. Moreover, the diagonal entries are all zero, because a paper can't cite itself.

We describe some important properties of the publication matrix and the citation matrix below.
\begin{itemize}
	\item The \emph{row sum} of the row $i, \sum_{j=1}^{n} m_{ij}$ is the number of papers published by author $a_i$.
	\item The \emph{column sum} of the column $j, \sum_{i=1}^{m} m_{ij}$ is the number of authors who have written the paper $p_j$.
	\item There is no self-citation by any paper, \emph{i.e.}, $c_{jj} = 0$ for $j=1,\ldots,n$.
	\item If the papers are numbered in \emph{decreasing order of time} (newest first), the resulting citation matrix will be \emph{upper-triangular}.
\end{itemize}

\begin{figure}[ht]
\begin{center}
\includegraphics[scale=0.6]{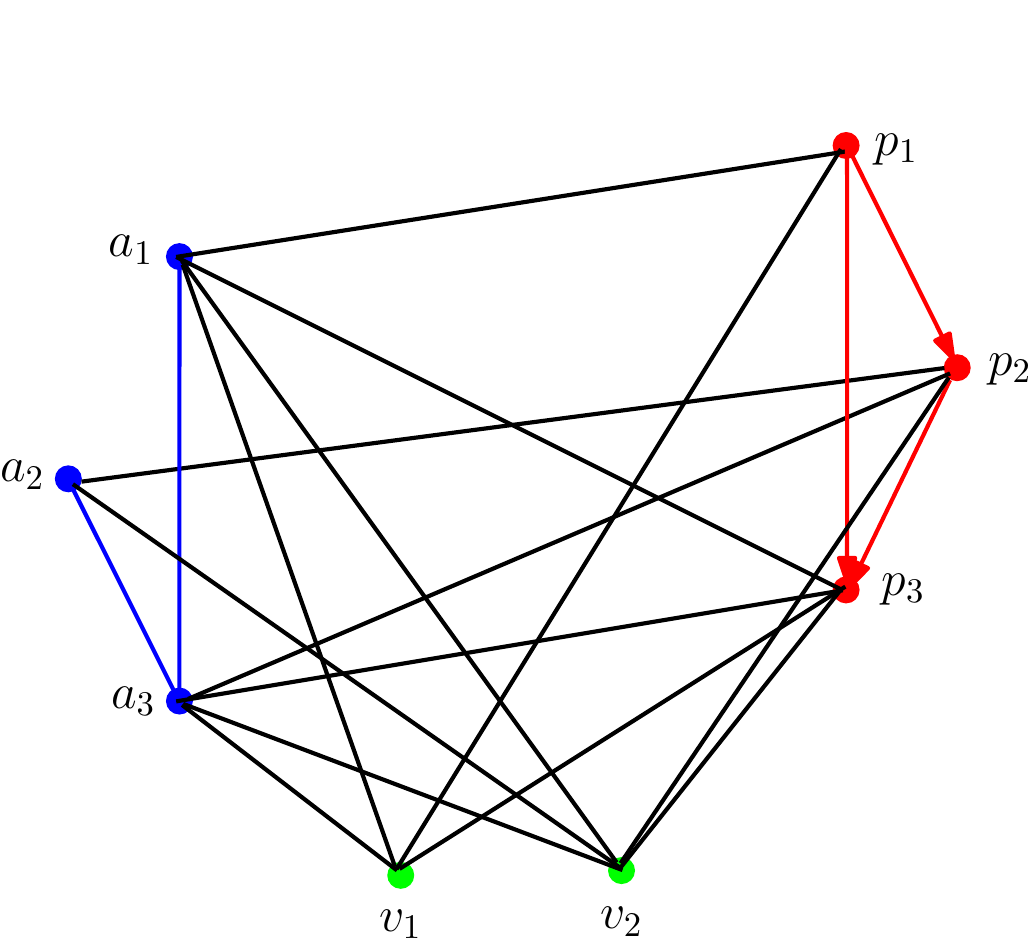}
\caption{The publication graph, collaboration graph and citation graph showing authors, papers and venues with their edges labeled in black, blue and red colors respectively.}
\label{publication-graph}
\end{center}
\end{figure}

An example of a publication graph, a collaboration graph and a citation graph is given in Figure \ref{publication-graph}.
 
The following sets are important for further development.
\begin{enumerate}
	\item For an author $a$, $AUTHORS(a)$ is defined as the set of co-authors of author $a$, \emph{i.e.,} the set of authors who have written at least one paper with $a$. In other words, $AUTHORS(a) = \{b \in \cA: (a,b) \in E_A\}$.
	\item For an author $a$, $PAPERS(a)$ is defined as the set of papers written by author $a$. In other words, $PAPERS(a) = \{p \in \cP: (a,p) \in E_P\}$.
	\item For an author $a$, $VENUES(a)$ is defined as the set of venues in which author $a$ has published. In other words, $VENUES(a) = \{v \in \cV: (a,v) \in E_P\}$.
	\item For a paper $p$, $AUTHORS(p)$ is defined as the set of authors who have written paper $p$. In other words, $AUTHORS(p) = \{a \in \cA: (a,p) \in E_P\}$.
	\item For a paper $p$, $CITE(p)$ is defined as the set of papers who have cited paper $p$. In other words, $CITE(p) = \{q \in \cP: (q,p) \in E_C\}$.
	\item For a paper $p$, $REF(p)$ is defined as the set of papers which have been given as reference (cited) by paper $p$. In other words, $REF(p) = \{q \in \cP: (p,q) \in E_C\}$.
	\item For a paper $p$, $VENUES(p)$ is defined as the set of venues in which paper $p$ was published. In other words, $VENUES(p) = \{v \in \cV: (p,v) \in E_P\}$. Note that, usually $VENUES(p)$ will have only one element $v_k$, where the paper $p$ was published. But it may happen that the same paper may be published in a workshop, conference and journal. So, we keep the definition general.
	\item For a venue $v$, $AUTHORS(v)$ is defined as the set of authors who have published in the venue $v$. In other words, $AUTHORS(v) = \{a \in \cA: (a,v) \in E_P\}$.
	\item For a venue $v$, $PAPERS(v)$ is defined as the set of papers published in the venue $v$. In other words, $PAPERS(v) = \{p \in \cP: (p,v) \in E_P\}$.
\end{enumerate}

\section{Our Algorithm}
Our goal is to assign scores to authors and papers using the structure of the publication and citation graphs, so that important authors and papers get higher scores.
\subsection{Computing Author, Paper and Venue Scores}
For each author $a_i$, we have an \emph{author score} (\emph{a-score}) $x_i$. For each paper $p_j$, we have a \emph{paper score} (\emph{p-score}) $y_j$. For each venue $v_k$, we have a \emph{venue score} (\emph{v-score}) $z_k$. We represent the set of author scores as a column vector $\xx = (x_1,\ldots,x_m)^T$, the set of paper scores as a column vector $\yy = (y_1,\ldots,y_n)^T$, and the set of venue scores as a column vector $\zz = (z_1,\ldots,z_r)^T$. We initialize all author, paper and venue scores to one, \emph{i.e.}, $\xx = \mathbf{1}_m, \yy = \mathbf{1}_n$, and $\zz = \mathbf{1}_r$. Then, we iteratively update the $a$-scores, $p$-scores and $v$-scores using the following rules.

\begin{enumerate}
	\item For each author $a_i$, its \emph{normalized $a$-score} $\bar{x}_i$ is given by the $a$-score $x_i$ divided by the sum of the number of co-authors, the number of papers written by him and the number of venues where his papers are published. In other words, $\bar{x}_i = \frac{x_i}{apv_i}$, for $i=1,\ldots,m$, where $apv_i = |AUTHORS(a_i)| + |PAPERS(a_i)| + |VENUES(a_i)|$ is the sum of the number of co-authors, the number of papers written and the number of venues where author $a_i$'s papers are published.
	\item For each paper $p_j$, its \emph{normalized $p$-score} $\bar{y}_j$ is given by the $p$-score $y_j$ divided by the sum of the number of authors who have written the paper, the number of papers who have cited the paper and the number of venues where the paper is published. In other words, $\bar{y}_j = \frac{y_j}{acv_j}$, for $j=1,\ldots,n$, where $acv_j = |AUTHORS(p_j)| + |CITE(p_j)| + |VENUES(p_j)|$ is the sum of the number of authors who have written the paper, the number of papers who have cited the paper and the number of venues where the paper is published.
	\item For each venue $v_k$, its \emph{normalized $v$-score} $\bar{z}_k$ is given by the $v$-score $z_k$ divided by the sum of the number of authors and papers published in venue $v_k$. In other words, $\bar{z}_k = \frac{z_k}{ap_k}$, for $k=1,\ldots,r$, where $ap_k = |AUTHORS(v_k)| + |PAPERS(v_k)|$ is the number of authors and papers published in venue $v_k$.
\end{enumerate}

The iterative update rules are given below.
\begin{align}
x_i &= \sum_{l \in AUTHORS(i)} \bar{x}_l + \sum_{j \in PAPERS(i)} \bar{y}_j + \sum_{k \in VENUES(i)} \bar{z}_k \\
y_j &= \sum_{i \in AUTHORS(j)} \bar{x}_i + \sum_{l \in CITE(j)} \bar{y}_l + \sum_{k \in VENUES(j)} \bar{z}_k \\
z_k &= \sum_{i \in AUTHORS(k)} \bar{x}_i + \sum_{j \in PAPERS(k)} \bar{y}_j
\end{align}

The solution of these system of recursive, simultaneous equations gives the author, paper and venue scores. We have proved some results on the convergence of these equations. We are calculating the scores for real graphs from \emph{Microsoft Academic Graph}, \emph{Google Scholar} and \emph{AMiner}. These results will be reported in the full version of this paper.

\section{Conclusion and Future Work}
Here are some future directions to work on.
\begin{itemize}
	\item Can we develop an axiomatic approach to citation analysis by starting from a set of basic axioms?
	\item In a real-world scenario, authors, papers and venues will be added over time. Dynamically modifying the scores from the current scores in an incremental manner is a challenging problem.
	\item Applying this framework to customer recommendation of products will be an important problem.
	\item This is an example of an \emph{interdependent graph}. How to extend our method to other interdependent graphs?
	\item Using other parameters such as time of publication and age of authors can be useful.
\end{itemize}

\bibliographystyle{abbrv}
\bibliography{citation}
%

\end{document}